\documentclass[twoside,reqno,11pt]{my_fcaa}

\usepackage{amsthm}
\usepackage{amsmath}
\usepackage{latexsym}
\usepackage{amsfonts}
\usepackage{amssymb}

\newtheoremstyle{defi}%
  {15pt}          %
  {15pt} %
  {\rm}  %
  {\parindent}    %
  {\sc}  %
  {. }   %
  { }    %
  {}     %
\theoremstyle{defi}
\newtheorem{definition}{Definition}[section]
\newtheorem{remark}{Remark}[section]

\usepackage{hyperref}

\usepackage{amsmath,amsbsy,amssymb,latexsym}
\usepackage{amssymb,amsthm}

\newcommand{\opab}{{^CD^{\alpha,\beta}_{\gamma}}}

\newcommand{\opabp}{{^CD^{{\alpha}_{1},{\beta}_{1}}_{{\gamma}_{1}}}}
\newcommand{\opabi}{{^CD^{\alpha_{i},\beta_{i}}_{\gamma_{i}}}}
\newcommand{\opabN}{{^CD^{\alpha_N,\beta_N}_{\gamma_N}}}

\newcommand{\opbac}{D^{\beta,\alpha}_{1-\gamma}}
\newcommand{\opbaci}{{D^{\beta_{i},\alpha_{i}}_{1-\gamma_{i}}}}

\newcommand{\caprb}{{^C_xD_b^\beta}}
\newcommand{\capla}{{^C_aD_x^\alpha}}

\title[Towards a combined fractional mechanics \dots]{Towards a combined fractional mechanics \\ [3pt] and quantization}

\author[A.B. Malinowska, D.F.M. Torres]{Agnieszka B. Malinowska $^1$, Delfim F. M. Torres $^2$}

\begin{document}

\begin{abstract}

A fractional Hamiltonian formalism is introduced for the recent
combined fractional calculus of variations. The Hamilton--Jacobi
partial differential equation is generalized to be applicable for
systems containing combined Caputo fractional derivatives. The
obtained results provide tools to carry out the quantization of
nonconservative problems through combined fractional canonical
equations of Hamilton type.

\bigskip


\smallskip


\end{abstract}


\maketitle


\section{Introduction}

The conservative physical systems imply frictionless motion and are
simplification of the real dynamical world. The systems in Nature
always contain internal damping and are subject to some external
forces that do not store energy and which are not equivalent to the
gradient of a potential. For nonconservative dynamical systems the
energy conservation law is broken and, as a consequence, the
standard Hamiltonian formalism is no longer valid for describing the
behavior of the system. Moreover, not only the energy, but also
other physical quantities such as linear or angular momentums, are
not conserved: Noether's classical theorem ceases to be valid
\cite{MyID:062}.

In order to model nonconservative dynamical systems, a novel
approach to variational calculus was proposed by Riewe in 1996 and
1997, \cite{rie,rie2}. The idea is based on the concept of
fractional differentiation and fractional integration,
\cite{ref:1,book:Baleanu}. Roughly speaking, the appearance of
fractional equations is natural due to physical reasons of
long-range dissipation, fractional wave propagation in complex media
and long memory. Fractional-order Euler--Lagrange equations and the
Hamilton formalism with fractional derivatives are investigated in
\cite{ref:4,ref:5}. For a fractional Hamiltonian analysis of
irregular systems, see \cite{ref:6}.

The fractional calculus of variations is nowadays an emerging field
--- see \cite{Shakoor:01,MyID:147,MyID:172,book:Baleanu,MyID:152,ref:3}
and references therein. A general fractional variational calculus,
based on a combined Caputo derivative, was recently introduced by
the authors in \cite{MyID:217} and further developed in
\cite{MyID:206} and \cite{MyID:207}. In \cite{ref:2}, a
generalization of the fractional variational calculus is given by
considering generalized Hilfer's fractional derivatives with two and
three parameters. Here we study the importance of the new combined
variational calculus in fractional mechanics, describing
nonconservative dynamic systems.

In Section~\ref{sec:2} we recall the combined fractional
Euler--Lagrange equations. Our results are then given in
Section~\ref{sec:3}. In Section~\ref{sec:3.1} we show that combined
fractional Hamiltonian equations of motion can be obtained for
nonconservative systems. The constants of motion cease to be valid
and a new notion is introduced in Section~\ref{sec:3.2}. Canonical
fractional transformations of first and second kind are studied in
Section~\ref{sub:sec:1st} and Section~\ref{sub:sec:2nd},
respectively. Subsequently, in Section~\ref{sec:3.5}, an
Hamilton--Jacobi type equation is derived and a fractional quantum
wave equation suggested.

The understanding of the Hamiltonian dynamical systems, classical or
quantum, has been a long standing theoretical question since Dirac's
quantization of the classical electromagnetic field. The combined
fractional Hamiltonian approach here introduced seems to be a
promising direction of research.


\section{Preliminaries}
\label{sec:2}

Let $\alpha, \beta \in(0,1)$ and $\gamma\in [0,1]$. The fractional
derivative operator $\opab$ was introduced in Malinowska and Torres
\cite{MyID:217}, as
$$\opab :=\gamma \, \capla + (1-\gamma) \,
\caprb.$$
 For $\gamma = 0$ and $\gamma = 1$ we obtain the standard
Caputo operators: ${^CD^{\alpha,\beta}_{0}} =\caprb$ and
${^CD^{\alpha,\beta}_{1}} =\capla$. Consider a fractional Lagrangian
\begin{equation}
\label{lag}
L\left(t,\mathbf{q},\opab \mathbf{q}\right)
\end{equation}
depending on time $t$,
the coordinates $\mathbf{q}=[q_1,\ldots, q_N]$ and their fractional
velocities $\opab \mathbf{q}=\left[\opabp q_1,\ldots,\opabN q_N\right]$.
The Euler--Lagrange equations corresponding to \eqref{lag}
were studied in \cite{MyID:206} and form the following system of $N$
fractional differential equations:
\begin{equation}
\label{E-L1}
\frac{\partial L}{\partial q_i}+\opbaci \frac{\partial L}{\partial \opabi q_i} = 0,
\quad i=1,\ldots N,
\end{equation}
where
$$\opbaci := (1-\gamma_i){_aD_x^{\beta_i}}+\gamma_i
{_xD_b^{\alpha_i}}$$
with ${_aD_x^{\beta_i}}$ and ${_xD_b^{\alpha_i}}$ denoting the classical
left and right Riemann--Liouville fractional derivatives.

\begin{remark} 
With the notation from \cite{MyID:206,MyID:207},
$$
\partial_{i+1}L=\frac{\partial L}{\partial q_i} \text{ and }
\partial_{N+1+i}L=\frac{\partial L}{\partial \opabi q_i}, \quad i=1,\ldots N.
$$
In contrast with \cite{MyID:206,MyID:207},
we adopt here the notation used in mechanics \cite{gold,rie2}.
\end{remark}

More general necessary optimality conditions than \eqref{E-L1} are
found in \cite{MyID:206,MyID:207}. However, all our previous works
\cite{MyID:217,MyID:206,MyID:207} on combined fractional derivatives
consider the Lagrangian approach only. In contrast, here we develop
the combined fractional Hamiltonian formalism. We show that
\eqref{E-L1} can be reduced to a special Hamiltonian system of $2N$
fractional differential equations: the combined canonical fractional
equations.


\section{Main Results}
\label{sec:3}

In analogy with the classical mechanics, let us introduce the
canonical momenta $p_i$ by
\begin{equation}
\label{mom}
p_i=\frac{\partial L}{\partial \opabi q_i}, \quad  i=1,\ldots N.
\end{equation}


\subsection{Combined Euler--Lagrange equations in Hamiltonian form}
\label{sec:3.1}

Assume that $\left|\frac{\partial(p_1,\ldots,p_N)}{\partial\left(
\opabp q_1,\ldots,\opabN q_N\right)}\right|\neq 0$. Then, by the
implicit function theorem, we can locally solve equations
\eqref{mom} with respect to \hfil \break $\opabp q_1$, \ldots,
$\opabN q_N$. The fractional Hamiltonian is defined by
\begin{equation}
\label{ham}
H(t,\mathbf{q},\mathbf{p})
=\sum_{i=1}^{N}p_i\opabi q_i-L(t,\mathbf{q},\opab \mathbf{q}),
\end{equation}
where $\opabi q_i$ are regarded as functions
of variables $t,q_1$, \ldots, $q_N,p_1$, \ldots, $p_N$.
Therefore, the differential of $H$ is given by
\begin{equation}
\label{dham}
dH = \sum_{i=1}^{N}\frac{\partial H}{\partial q_i}dq_i
+\sum_{i=1}^{N}\frac{\partial H}{\partial p_i}dp_i
+\frac{\partial H}{\partial t}dt.
\end{equation}
From the defining equality \eqref{ham}, we can also write
\begin{multline}
\label{dham1}
dH=\sum_{i=1}^{N} {^CD^{\alpha_{i},\beta_{i}}_{\gamma_{i}}} q_i dp_i
+\sum_{i=1}^{N}p_i d {^CD^{\alpha_{i},\beta_{i}}_{\gamma_{i}}} q_i
-\sum_{i=1}^{N}\frac{\partial L}{\partial q_i}dq_i\\
-\sum_{i=1}^{N}\frac{\partial L}{\partial
{^CD^{\alpha_{i},\beta_{i}}_{\gamma_{i}}} q_i}
d {^CD^{\alpha_{i},\beta_{i}}_{\gamma_{i}}} q_i
-\frac{\partial L}{\partial t}dt.
\end{multline}
The terms containing $d\opabi q_i$ in \eqref{dham1} cancel
because of the definition of canonical momenta.
Applying relations \eqref{E-L1}, we get
\begin{equation*}
dH=\sum_{i=1}^{N} {^CD^{\alpha_{i},\beta_{i}}_{\gamma_{i}}} q_i dp_i
+\sum_{i=1}^{N}{D^{\beta_{i},\alpha_{i}}_{1-\gamma_{i}}}p_idq_i
-\frac{\partial L}{\partial t}dt.
\end{equation*}
Comparison with \eqref{dham} furnishes the following set of $2N$ relations:
\begin{equation}
\label{e_h}
\frac{\partial H}{\partial p_i}
= \opabi q_i, \quad \frac{\partial H}{\partial q_i}
= \opbaci p_i,\quad i=1,\ldots,N,
\end{equation}
which we can call the \emph{combined fractional canonical equations of Hamilton}.
They constitute a set of $2N$ fractional order equations of motion replacing the
Euler--Lagrange equations \eqref{E-L1}. Moreover,
\begin{equation}
\label{e_h_1}
\frac{\partial H}{\partial t}=- \frac{\partial L}{\partial t}.
\end{equation}
For integer-order derivatives, it can be shown that
\begin{equation*}
\frac{d H}{d t}=\frac{\partial H}{\partial t}
= - \frac{\partial L}{\partial t}
\end{equation*}
(see, \textrm{e.g.}, \cite[p.~220]{gold}). Hence, if $L$ (and, in
consequence of \eqref{e_h_1}, also $H$) is not an explicit function
of $t$, \textrm{i.e.}, in the autonomous case, then $H$ is a
constant of motion. This is a well-known consequence of Noether's
theorem \cite{MyID:012,MyID:032}. With noninteger-order derivatives
this is not the case, see \cite{MyID:089,MyID:149}. Observe that
using equations \eqref{e_h}, we can write
\begin{equation*}
\frac{d H}{d t}=\sum_{i=1}^{N} \left({^CD^{\alpha_{i},\beta_{i}}_{\gamma_{i}}} q_i \frac{dp_i}{dt}
+\opbaci p_i \frac{dq_i}{dt}\right)+\frac{\partial H}{\partial t}.
\end{equation*}
In contrast with the integer-order case, the terms containing fractional and classical derivatives
of coordinates and momenta do not cancel. Therefore, in general we obtain nonconservative systems
and classical constants of motion cease to be valid \cite{MyID:062,rie}.
To deal with the problem we introduce the notion of fractional constant of motion.


\subsection{Fractional constants of motion}
\label{sec:3.2}

To account for the presence of dissipative terms, we propose
the following definition of fractional constant of motion.

\begin{definition}
We say that a function $C(t,\mathbf{q},\mathbf{p})$ is a fractional
constant of motion of order $(\alpha,\beta,\gamma)$ if
$\opbac \left[t \mapsto C\left(t,\mathbf{q}(t),\mathbf{p}(t)\right)\right]$
is the null function along any pair $(\mathbf{q},\mathbf{p})$
satisfying the $2N$ combined fractional canonical equations \eqref{e_h}.
\end{definition}

It follows from \eqref{e_h} that if $q_i$ is absent in the fractional Hamiltonian,
then $\opbaci p_i = 0$, \textrm{i.e.}, if $\frac{\partial L}{\partial q_i} = 0$, then
$p_i=\frac{\partial L}{\partial \opabi q_i}$ is a fractional constant of motion of order
$(\alpha_i,\beta_i,\gamma_i)$.


\subsection{Canonical fractional transformations of first kind}
\label{sub:sec:1st}

We now look for transformations under which the combined fractional canonical
equations of Hamilton \eqref{e_h} preserve their canonical form.
Consider the simultaneous transformation of independent coordinates
and momenta, $q_i$ and $p_i$, to a new set $Q_i$ and $P_i$
with transformation equations
\begin{equation*}
Q_i=Q_i(t,\mathbf{q},\mathbf{p}),\quad
P_i=P_i(t,\mathbf{q},\mathbf{p}), \quad
i=1,\ldots,N.
\end{equation*}
The new $Q_i$ and $P_i$ are canonical coordinates provided there exists
some function $K(t,\mathbf{Q},\mathbf{P})$ such that the equations of motion
in the new set are in Hamiltonian form:
\begin{equation}
\label{e_h_n}
\frac{\partial K}{\partial P_i} = \opabi Q_i,
\quad \frac{\partial K}{\partial Q_i}
= \opbaci P_i,\quad i=1,\ldots,N.
\end{equation}
Transformations for which equations \eqref{e_h_n} are valid are said to be canonical.
Function $K$ plays the role of the Hamiltonian in the new coordinates set.
If $Q_i$ and $P_i$ are to be canonical coordinates, then they must satisfy
the modified fractional Hamiltonian principle of form
\begin{equation}
\label{prin_old}
\delta\int_{a}^{b}\left(\sum_{i=1}^{N}P_i\opabi Q_i
-K(t,\mathbf{Q},\mathbf{P})\right)dt = 0.
\end{equation}
At the same time, the original coordinates satisfy the similar principle
\begin{equation}
\label{prin_new}
\delta\int_{a}^{b}\left(\sum_{i=1}^{N}p_i\opabi
q_i-H(t,\mathbf{q},\mathbf{p})\right)dt = 0.
\end{equation}
The simultaneous validity of equations \eqref{prin_old} and \eqref{prin_new}
means that integrands of the two integrals can differ at most by a total derivative
of an arbitrary function $F$, sometimes called a gauge term \cite{MyID:025}:
\begin{equation*}
\frac{d}{dt}F+\sum_{i=1}^{N}P_i\opabi Q_i
-K(t,\mathbf{Q},\mathbf{P})
=\sum_{i=1}^{N}p_i\opabi q_i
-H(t,\mathbf{q},\mathbf{p}).
\end{equation*}
The arbitrary function $F$ works as a generating function of the transformation,
and is very useful in developing direct methods to solve problems of the calculus
of variations \cite{MyID:154,MyID:183} and optimal control \cite{MyID:077,MyID:096}.
Function $F$ is only specified up to an additive constant.
In order to produce transformations between the two sets of canonical variables,
$F$ must be a function of both old and new variables.
For mechanics involving fractional derivatives (\textrm{cf.} \cite{rie})
we need to introduce variables $\bar{q}_i$ and $\bar{Q}_i$, $i=1,\ldots,N$, satisfying
\begin{equation*}
\frac{d \bar{q}_i}{dt}=\opabi q_i,\quad  \frac{d \bar{Q}_i}{dt}=\opabi Q_i.
\end{equation*}
For $\gamma_i=1$, $\alpha_i=1$, $\beta_i=0$, $i=1,\ldots,N$,
these new coordinates are the same as the usual canonical coordinates.
However, when dealing with fractional derivatives, the coordinates $\bar{q}_i$
and $\bar{Q}_i$ will not be canonical, so all canonical expressions
must be written in terms of $\opabi q_i$ and $\opabi Q_i$.
For a generating function $F_1\left(t,\bar{\mathbf{q}},\bar{\mathbf{Q}}\right)$,
the integrands of \eqref{prin_old} and \eqref{prin_new} are connected by the relation
\begin{equation}
\label{gener1}
\frac{d}{dt} F_1(t,\bar{\mathbf{q}}, \bar{\mathbf{Q}})
=\sum_{i=1}^{N}p_i\opabi q_i-H(t,\mathbf{q},\mathbf{p})
-\sum_{i=1}^{N}P_i\opabi Q_i+K(t,\mathbf{Q},\mathbf{P}).
\end{equation}
Because
\begin{equation*}
d F_1=\sum_{i=1}^{N}\frac{\partial F_1}{\partial\bar{q}_i}d\bar{q}_i
+ \sum_{i=1}^{N}P_i \frac{\partial F_1}{\partial\bar{Q}_i}d\bar{p}_i
+\frac{\partial F_1}{\partial t}dt,
\end{equation*}
we obtain
\begin{equation}
\label{trans1}
\frac{\partial F_1}{\partial t}=K-H,\quad
p_i=\frac{\partial F_1}{\partial\bar{q}_i},\quad
-P_i=\frac{\partial F_1}{\partial\bar{Q}_i},
\end{equation}
$i=1,\ldots,N$. We call to \eqref{trans1}
the canonical fractional transformations of first kind.


\subsection{Canonical fractional transformations of second kind}
\label{sub:sec:2nd}

We shall now introduce canonical transformations of second type
with a generating function $F_2(t,\bar{\mathbf{q}},\mathbf{P})$.
We begin by noting that the transformation from  $\bar{q}_i$ and $\bar{Q}_i$
to  $\bar{q}_i$ and $P_i$ can be accomplished by a Legendre transformation.
Indeed, by equations \eqref{trans1}, $-P_i=\frac{\partial F_1}{\partial\bar{Q}_i}$.
This suggests that the generating function $F_2$ can be defined by
\begin{equation*}
F_2(t,\bar{\mathbf{q}},\mathbf{P})
=F_1(t,\bar{\mathbf{q}}, \bar{\mathbf{Q}})
+\sum_{i=1}^{N}P_i\bar{Q}_i.
\end{equation*}
From equation \eqref{gener1} we obtain
\begin{equation*}
\begin{split}
\sum_{i=1}^{N}p_i\frac{d \bar{q}_i}{dt}-H(t,\mathbf{q},\mathbf{p})
&-\sum_{i=1}^{N}P_i\frac{d \bar{Q}_i}{dt} +K(t,\mathbf{Q},\mathbf{P})\\
&=\frac{d}{dt}\left(F_2(t,\bar{\mathbf{q}},\mathbf{P})-\sum_{i=1}^{N}P_i\bar{Q}_i\right)\\
&=\frac{d}{dt}F_2(t,\bar{\mathbf{q}},\mathbf{P})-\sum_{i=1}^{N}P_i\frac{d\bar{Q}_i}{dt}
-\sum_{i=1}^{N}\bar{Q}_i\frac{dP_i}{dt}.
\end{split}
\end{equation*}
Hence,
\begin{equation*}
\left(\sum_{i=1}^{N}p_i\frac{d \bar{q}_i}{dt}-H\right)dt
+\left(\sum_{i=1}^{N}\bar{Q}_i\frac{dP_i}{dt} + K\right)dt=
d F_2(t,\bar{\mathbf{q}},\mathbf{P}).
\end{equation*}
Repeating the procedure followed for $F_1$ in Section~\ref{sub:sec:1st},
we obtain the transformation equations
\begin{equation}
\label{trans2}
\frac{\partial F_2}{\partial t}=K-H,\quad p_i
=\frac{\partial F_2}{\partial\bar{q}_i},\quad \bar{Q}_i
=\frac{\partial F_2}{\partial P_i}, \quad i=1,\ldots,N.
\end{equation}


\subsection{Fractional Hamilton--Jacobi equation}
\label{sec:3.5}

Under the assumption that $K$ is identically zero,
for integer order derivatives we know,
from equations \eqref{e_h_n},
that the new coordinates are constant.
The same situation occurs, under appropriate assumptions,
in the fractional setting
when in presence of Riemann--Liouville
or Caputo derivatives \cite{Golman,Rabei:HJ}.
In our general combined setting this is not the case,
in particular for the  new momentum $P$.
To solve the problem we proceed as follows.
Assume that the transformed Hamiltonian, $K$, is identically zero.
Then the equations of motion are
\begin{equation*}
\frac{\partial K}{\partial P_i}= \opabi Q_i=0,
\quad \frac{\partial K}{\partial Q_i} = \opbaci P_i=0,
\quad i=1,\ldots,N.
\end{equation*}
The function $K$ is related with the old Hamiltonian $H$ and the
generating function $F$ by $K=H+\frac{\partial F}{\partial t}$.
Hence, $K$ will be zero if, and only if, $F$ satisfies the equation
$H(t,\mathbf{q},\mathbf{p})+\frac{\partial F}{\partial t}=0$. It is
convenient to take $F$ as a function of $\bar{q}_i$ and $P_i$. Then,
from equations \eqref{trans2}, we can write
\begin{equation}
\label{H_J}
H\left(t,\mathbf{q}, \frac{\partial F_2}{\partial \bar{\mathbf{q}}}\right)
+\frac{\partial F_2}{\partial t}=0.
\end{equation}
Equation \eqref{H_J} is a fractional version of the Hamilton--Jacobi equation.
Therefore, the quantum wave equation with the combined $\opab$ fractional derivative
is suggested to be
\begin{equation*}
H\left(t,\mathbf{q},-i\hbar \frac{\partial}{\partial \bar{\mathbf{q}}}\right) \psi
=i\hbar\frac{\partial \psi}{\partial t},
\end{equation*}
where $\psi$ is a wave equation.


\section{Conclusion}

In this paper we give an Hamiltonian formulation to the general
fractional Euler--Lagrange equations obtained in \cite{MyID:217}.
Motivated by the recent results in the literature of Physics dealing
with fractional mechanics and nonconservative systems
\cite{abreu,Golman,Rabei:HJ}, we illustrate some possible
applications of our combined variational calculus in mechanics and
quantization.


\section*{Acknowledgments}

This work was supported by {\it FEDER} funds through {\it COMPETE}
--- Operational Programme Factors of Competitiveness ---
and by Portuguese funds through the {\it Center for Research and
Development in Mathematics and Applications} (University of Aveiro)
and the Portuguese Foundation for Science and Technology (FCT),
within project PEst-C/MAT/UI4106/2011 with COMPETE number
FCOMP-01-0124-FEDER-022690. A.B. Malinowska was also supported by
BUT grant S/WI/2/2011; and D.F.M. Torres by the FCT project
PTDC/MAT/113470/2009.



\bigskip \medskip

\it

\noindent
$^1$ Faculty of Computer Science \\
Bia{\l}ystok University of Technology \\
15-351 Bia\l ystok, POLAND  \\[4pt]
e-mail: a.malinowska@pb.edu.pl
\hfill Received: February 21, 2012 \\[15pt]
$^2$ Center for Research and Development in Mathematics and Applications \\
Department of Mathematics, University of Aveiro \\
3810-193 Aveiro, PORTUGAL \\[4pt]
e-mail: delfim@ua.pt


\end{document}